\documentclass[12pt,preprint]{aastex}
\def\ltsima{$\; \buildrel < \over \sim \;$}
\def\gtsima{$\; \buildrel > \over \sim \;$}
\def\lsim{\lower.5ex\hbox{\ltsima}}
\def\gsim{\lower.5ex\hbox{\gtsima}}
\def\lapp{\ifmmode\stackrel{<}{_{\sim}}\else$\stackrel{<}{_{\sim}}$\fi}
\def\gapp{\ifmmode\stackrel{>}{_{\sim}}\else$\stackrel{<}{_{\sim}}$\fi}

\newdimen\minuswidth    
\setbox0=\hbox{$-$}
\minuswidth=\wd0
\catcode`@=\active
\def@{\kern\minuswidth}
\setbox0=\hbox{\rm0}
 
\shorttitle{Blue Stragglers in M5}
\shortauthors{Lanzoni et al.}
 
\begin{document} 
 
\title{The Blue Straggler Population of the Globular Cluster M5
\footnote{Based on observations with the NASA/ESA HST, obtained at the Space
Telescope Science Institute, which is operated by AURA, Inc., under NASA
contract NAS5-26555. Also based on WFI observations collected at the European
Southern Observatory, La Silla, Chile, within the observing programs
62.L-0354 and 64.L-0439.}  }

\author{
B. Lanzoni\altaffilmark{1,2},
E. Dalessandro\altaffilmark{1,2},
F.R. Ferraro\altaffilmark{1},
C. Mancini\altaffilmark{3},
G. Beccari\altaffilmark{2,4,5},
R.T. Rood\altaffilmark{6},
M. Mapelli\altaffilmark{7},
S. Sigurdsson\altaffilmark{8}
}
\affil{\altaffilmark{1} Dipartimento di Astronomia, Universit\`a degli Studi
di Bologna, via Ranzani 1, I--40127 Bologna, Italy}
\affil{\altaffilmark{2} INAF--Osservatorio Astronomico di Bologna, via
Ranzani 1, I--40127 Bologna, Italy}  
\affil{\altaffilmark{3} Dipartimento di Astronomia e Scienza dello Spazio,
Universit\`a degli Studi di Firenze, Largo Enrico Fermi 2, I-- 50125
Firenze, Italy}
\affil{\altaffilmark{4} Dipartimento di Scienze della Comunicazione, 
Universit\`a degli Studi di Teramo, Italy}
\affil{\altaffilmark{5} INAF--Osservatorio Astronomico 
di Collurania, Via Mentore Maggini, I--64100 Teramo, Italy}
\affil{\altaffilmark{6} Department of Astronomy and Astrophysics, The
Pennsylvania State University, 525 Davey Lab, University Park, PA~16802}
\affil{\altaffilmark{7} S.I.S.S.A., Via Beirut 2 - 4, I--34014 Trieste, Italy}
\affil{\altaffilmark{8} Astronomy Department, University of Virginia,
P.O. Box 400325, Charlottesville, VA, 22904}

\date{20 March, 07}

\keywords{Globular clusters: individual (M5); stars: evolution -- binaries:
general - blue stragglers}

\begin{abstract}
By combining high-resolution HST and wide-field ground based observations, in
ultraviolet and optical bands, we study the Blue Stragglers Star (BSS)
population of the galactic globular cluster M5 (NGC 5904) from its very
central regions up to its periphery.  The BSS distribution is highly peaked
in the cluster center, decreases at intermediate radii and rises again
outward. Such a bimodal distribution is similar to those previously observed
in other globular clusters (M3, 47~Tucanae, NGC~6752).  As for these
clusters, dynamical simulations suggest that, while the majority of BSS in M5
could be originated by stellar collisions, a significant fraction (20-40\%)
of BSS generated by mass transfer processes in primordial binaries is
required to reproduce the observed radial distribution.  A candidate BSS has
been detected beyond the cluster tidal radius.  If confirmed, this could
represent an interesting case of an "evaporating" BSS.
\end{abstract}

\section{INTRODUCTION}

In globular cluster (GC) color-magnitude diagrams (CMD) blue straggler stars
(BSS) appear to be brighter and bluer than the Turn-Off (TO) stars and lie
along an extension of the Main Sequence.  Since BSS mimic a rejuvenated
stellar population with masses larger than the normal cluster stars (this is
also confirmed by direct mass measurements; e.g. Shara et al. 1997), they are
thought to be objects that have increased their initial mass during their
evolution by means of some process.  Two main scenarios have been proposed
for their formation: the {\it collisional scenario} suggests that BSS are the
end-products of stellar mergers induced by collisions (COL-BSS), while in the
\emph{mass-transfer} scenario BSS form by the mass-transfer activity between
two companions in a binary system (MT-BSS), possibly up to the complete
coalescence of the two stars.  Hence, understanding the origin of BSS in
stellar clusters provides valuable insight both on the binary evolution
processes and on the effects of dynamical interactions on the (otherwise
normal) stellar evolution.

The relative efficiency of the two formation mechanisms is thought to depend
on the environment \citep{fp92, fe99a, bel02, fe03}.  COL-BSS are expected to
be formed preferentially in high-density environments (i.e., the GC central
regions), where stellar collisions are most probable, and MT-BSS should
mainly populate lower density environments (the cluster peripheries), where
binary systems can more easily evolve in isolation without suffering
exchanges or ionization due to gravitational encounters.  The overall
scenario is complicated by the fact that primordial binaries can also sink to
the core due to mass segregation processes, and ``new'' binaries can be
formed in the cluster centers by gravitational encounters. The two formation
mechanisms are likely to be at work simultaneously in every GC (see the case
of M3 as an example; Ferraro et al. 1993, 1997), but the identification of
the cluster properties that mainly affect their relative efficiency is still
an open issue.

One possibility for distinguishing between the two types of BSS is offered by
high-resolution spectroscopic studies. Anomalous chemical abundances are
expected at the surface of BSS resulting from MT activity \citep{sarna96},
while they are not predicted in case of a collisional formation
\citep{lomb95}.  Such studies have just become feasible, and the results
found in the case of 47~Tucanae \citep[47~Tuc;][]{fe06a} are encouraging.
The detection of unexpected properties of stars along standard
evolutionary sequences (e.g., variability, anomalous population
fractions, or peculiar radial distributions) can help estimating the
fraction of binaries within a cluster \citep[see, e.g.,][]{bail94,
albr01, bel02, bec06}, but such evidence does not directly allow the
determination of the relative efficiency of the two BSS formation
processes.

The most widely applicable tool to probe the origin of BSS is their radial
distribution within the clusters \citep[see][for a review]{fe06}. This has
been observed to be \emph{bimodal} (i.e., highly peaked in the cluster
centers and peripheries, and significantly lower at intermediate radii) in at
least 4 GCs: M3 \citep{fe97}, 47~Tuc \citep{fe04}, NGC~6752 \citep{sab04},
and M5 \citep[][hereafter W06]{w06}. Preliminary evidence of bimodality has
also been found in M55 \citep{zag97}.  Dynamical simulations suggest that the
bimodal radial distributions observed in M3, 47~Tuc and NGC~6752 \citep{ma04,
ma06} result from $\sim 40-50\%$ of MT-BSS with the balance being COL-BSS.
In this context, the case of $\omega~$Cen is atypical: the BSS radial
distribution in this cluster is flat \citep{fe06b}, and mass segregation
processes have not yet played a major role, thus implying that this system is
populated by a vast majority of MT-BSS \citep{ma06}.  These results
demonstrate that detailed studies of the BSS radial distribution within GCs
are very powerful tools for better understanding the complex interplay
between dynamics and stellar evolution in dense stellar systems.

In the present paper we extend this kind of investigation to M5 (NGC 5904).
With HST-WFPC2 and -ACS ultraviolet and optical high-resolution images of the
core we have been able to efficiently detect the BSS population even in the
severely crowded central regions.  Moreover, with wide-field optical
observations performed with ESO-WFI we sampled the entire cluster extension.
The combination of these two data sets allowed us to study the dynamical
properties of M5, accurately redetermining its center of gravity, its surface
density profile, and the BSS radial distribution over the entire cluster.
The BSS population of M5 has been recently studied by W06, but we have
extended the analysis to larger distances from the cluster center, and we
have used Monte-Carlo dynamical simulations to interpret the observational
results.

\section{OBSERVATIONS AND DATA ANALYSIS}

\subsection{The data sets}
The present study is based on a combination of two different
photometric data sets:

\emph{1. The high-resolution set} -- It consists of a series of ultraviolet
(UV) and optical images of the cluster center obtained with HST-WFPC2
(Prop. 6607, P.I. Ferraro).  To efficiently resolve the stars in the highly
crowded central regions, the Planetary Camera (PC, being the highest
resolution instrument: $0\farcs 046/$pixel) has been pointed approximately on
the cluster center, while the three Wide Field Cameras (WF, having a lower
resolution: $0\farcs 1/$pixel) have been used to sample the surrounding
regions.  Observations have been performed through filter F255W (medium UV)
in order to efficiently select the BSS and horizontal branch (HB)
populations, and through filters F336W (approximately corresponding to an $U$
filter) and F555W ($V$) for the red giant branch (RGB) population and to
guarantee a proper combination with the ground-based data set (see below).
The photometric reduction of the high-resolution images was carried out using
ROMAFOT (Buonanno et al. 1983), a package developed to perform accurate
photometry in crowded fields and specifically optimized to handle
under-sampled Point Spread Functions (PSFs; Buonanno $\&$ Iannicola 1989), as
in the case of the HST-WF chips.

To obtain a better coverage of the innermost regions of the cluster, we have
also used a set of public HST-WFPC2 and HST-ACS observations.  The HST-WFPC2
data set has been obtained through filters F439W ($B$) and F555W ($V$) by
\citet{pio02}, and because of the different orientation of the camera, it is
complementary to ours. Additional HST-ACS data in filters F435W ($B$), F606W
($V$), and F814W ($I$) have been retrieved from the ESO-STECF Science
Archive, and have been used to sample the central area not covered by the
WFPC2 observations.  All the ACS images were properly corrected for geometric
distortions and effective flux (over the pixel area) following the
prescriptions of \citet{si05}.  The photometric analysis was performed
independently in the three drizzled images by using the aperture photometry
code SExtractor \citep[{\it Source-Extractor};] []{be96}, and adopting a
fixed aperture radius of 2.5 pixels ($0.125\arcsec$). The magnitude lists
were finally cross-correlated in order to obtain a combined catalog.  The
adopted combination of the three HST data sets is sketched in Figure
\ref{fig:HST} and provided a good coverage of the cluster up to $r
=115\arcsec$.

\emph{2. The wide-field set} - A complementary set of wide-field $B$ and $V$
images was secured by using the Wide Field Imager (WFI) at the 2.2m ESO-MPI
telescope during an observing run in April 2000. 
Thanks to the exceptional imaging capabilities of WFI (each image consists of
a mosaic of 8 CCDs, for a global field of view of $34\arcmin\times
34\arcmin$), these data cover the entire cluster extension (see Figure
\ref{fig:WFI}, where the cluster is roughly centered on CCD $\# 7$).
The raw WFI images were corrected for bias and flat field, and the overscan
regions were trimmed using IRAF\footnote{IRAF is distributed by the National
Optical Astronomy Observatory, which is operated by the Association of
Universities for Research in Astronomy, Inc., under cooperative agreement
with the National Science Foundation.}  tools. The PSF fitting procedure was
performed independently on each image using DoPhot \citep{dophot}.  All
the uncertain detections, usually caused by photometric blends, stars near
the CCD gaps or saturated stars, have been checked one by one using ROMAFOT
(Buonanno et al. 1983).

\subsection{Astrometry and center of gravity}
The HST+WFI catalog has been placed on the absolute astrometric system by
adopting the procedure already described in Ferraro et al. (2001, 2003).  The
new astrometric Guide Star Catalog (GSC-II\footnote{Available at {\tt
http://www-gsss.stsci.edu/Catalogs/GSC/GSC2/GSC2.htm}.}) was used to search
for astrometric standard stars in the WFI field of view (FoV), and a
cross-correlation tool specifically developed at the Bologna Observatory
(Montegriffo et al. 2003, private communication) has been employed to obtain
an astrometric solution for each of the 8 CCDs.  Several hundred GSC-II
reference stars were found in each chip, thus allowing an accurate absolute
positioning of the stars.  Then, a few hundred stars in common between the
WFI and the HST FoVs have been used as secondary standards to place the HST
catalog on the same absolute astrometric system.  At the end of the procedure
the global uncertainties in the astrometric solution are of the order of $\sim
0\farcs 2$, both in right ascension ($\alpha$) and declination ($\delta$).

Given the absolute positions of individual stars in the innermost regions of
the cluster, the center of gravity $C_{\rm grav}$ has been determined by
averaging coordinates $\alpha$ and $\delta$ of all stars lying in the PC FoV
following the iterative procedure described in Montegriffo et al. (1995; see
also Ferraro et al. 2003, 2004).  In order to correct for spurious effects
due to incompleteness in the very inner regions of the cluster, we considered
two samples with different limiting magnitudes ($m_{555}<19.5$ and
$m_{555}<20$), and we computed the barycenter of stars for each sample.  The
two estimates agree within $\sim 1\arcsec$, giving $C_{\rm grav}$ at
$\alpha({\rm J2000}) = 15^{\rm h}\, 18^{\rm m}\, 33\fs 53$, $\delta ({\rm
J2000})= +2^{\rm o}\, 4\arcmin\, 57\farcs 06$, with a 1$\sigma$ uncertainty
of $0\farcs 5$ in both $\alpha$ and $\delta$, corresponding to about 10
pixels in the PC image. This value of $C_{\rm grav}$ is located at $\sim
4\arcsec$ south-west ($\Delta\alpha = -4\arcsec$, $\Delta\delta=-0\farcs 9$)
from that previously derived by \citet{har96} on the basis of the surface
brightness distribution.

\subsection{Photometric calibration and definition of the catalogs}

The optical HST magnitudes (i.e., those obtained through the WFPC2 filters
F439W and F555W, and through ACS filters F435W, F606W, F814W), as well as the
WFI $B$ and $V$ magnitudes have been all calibrated on the catalog of
\citet{sand96}.  The UV magnitudes $m_{160}$ and $m_{255}$ have been
calibrated to the \citet{holtz95} zero-points following \citet[][2001]{fe97},
while the U magnitude $m_{336}$ has been calibrated to \citet{dolph00}.

\begin{figure}[!p]
\begin{center}
\includegraphics[scale=0.7]{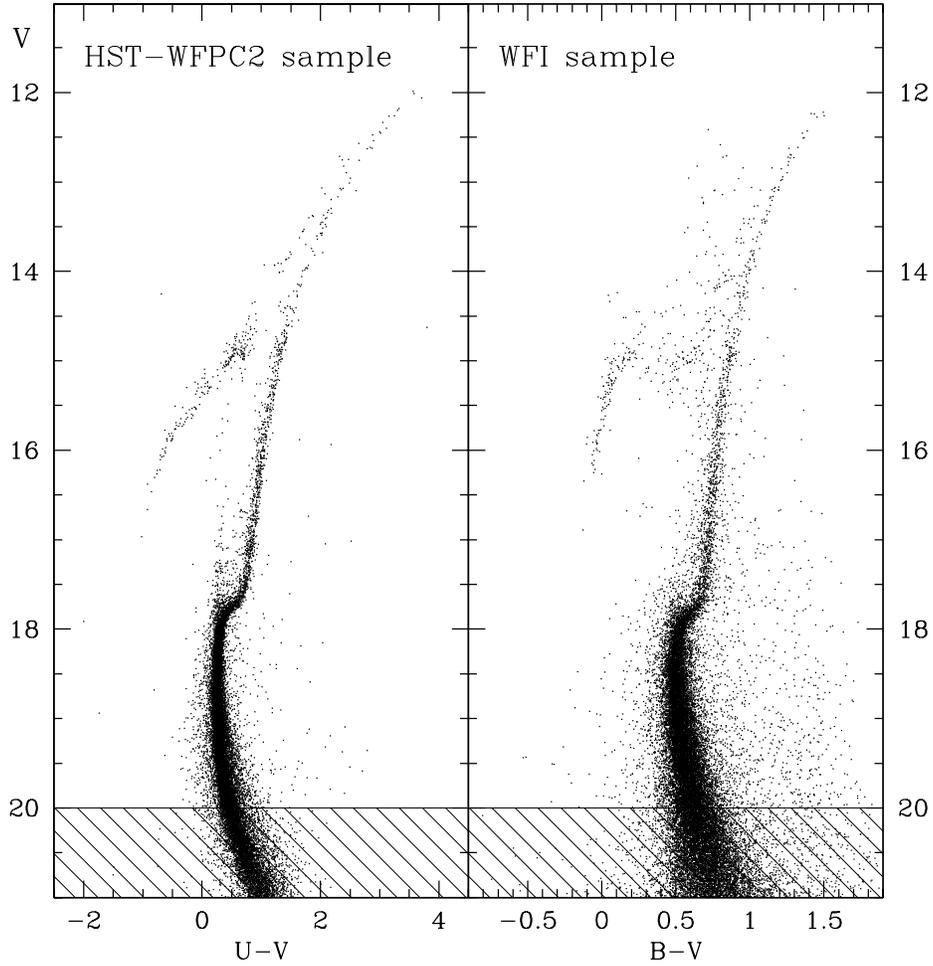}
\caption{Optical CMDs of the WFPC2-HST and the WFI samples. The hatched
regions indicate the magnitude limit ($V\le 20$) adopted for selecting the
stars used to construct the cluster surface density profile.}
\label{fig:CMD}
\end{center}
\end{figure}

In order to reduce spurious effects due to the low resolution of the
ground-based observations in the most crowded regions of the cluster, we use
only the HST data for the inner $115\arcsec$, this value being imposed by the
FoV of the WFPC2 and ACS cameras (see Figure \ref{fig:HST}).  In particular,
we define as \emph{HST sample} the ensemble of all the stars in the WFPC2 and
ACS combined catalog having $r\le115\arcsec$ from the center, and as
\emph{WFI sample} all stars detected with WFI at $r>115\arcsec$ (see Figure
\ref{fig:WFI}).  The CMDs of the HST and WFI samples in the $(V,~U-V)$ and
$(V,~B-V)$ planes are shown in Figure \ref{fig:CMD}.

\subsection{Density profile}
\label{sec:dens_prof}
We have determined the projected density profile over the entire cluster
extension, from $C_{\rm grav}$ out to $\sim 1400\arcsec\sim23\farcm 3$, by
direct star counts, considering only stars brighter than $V= 20$ (see Figure
\ref{fig:CMD}) in order to avoid incompleteness biases.  The brightest RGB
stars that are strongly saturated in the ACS data set have been excluded from
the analysis, but since they are few in number, the effect on the resulting
density profile is completely negligible.  Following the procedure already
described in Ferraro et al. (1999a, 2004), we have divided the entire HST+WFI
sample in 27 concentric annuli, each centered on $C_{\rm grav}$ and split in
an adequate number of sub-sectors. The number of stars lying within each
sub-sector was counted, and the star density was obtained by dividing these
values by the corresponding sub-sector areas.  The stellar density in each
annulus was then obtained as the average of the sub-sector densities, and its
standard deviation was estimated from the variance among the sub-sectors.

\begin{figure}[!p]
\begin{center}
\includegraphics[scale=0.7]{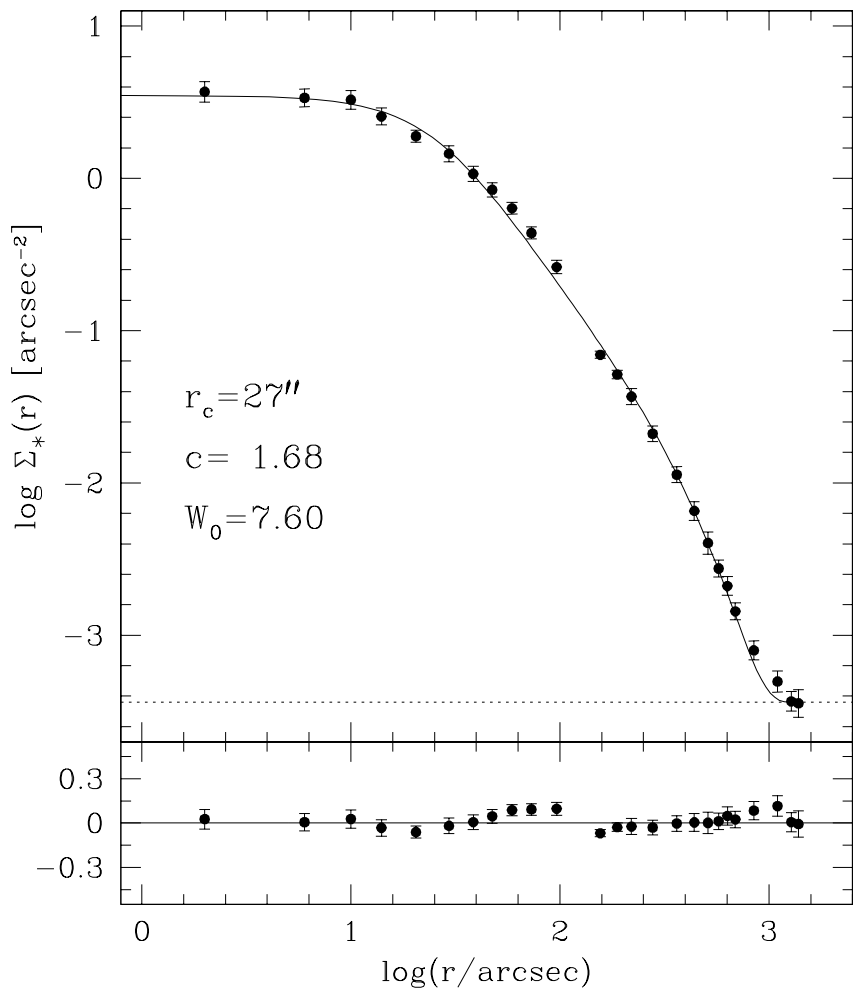}
\caption{Observed surface density profile (dots and error bars) and best-fit
King model (solid line). The radial profile is in units of number of stars
per square arcseconds.  The dotted line indicates the adopted level of the
background, and the model characteristic parameters (core radius $r_c$,
concentration $c$, dimensionless central potential $W_0$) are marked in the
figure. The lower panel shows the residuals between the observations and the
fitted profile at each radial coordinate.}
\label{fig:prof}
\end{center}
\end{figure}

The radial density profile thus derived is plotted in Figure \ref{fig:prof},
where we also show the best-fit mono-mass King model and the corresponding
values of the core radius and concentration: $r_c=27\arcsec$ (with a typical
error of $\sim \pm 2\arcsec$) and $c=1.68$, respectively.  These values
confirm that M5 has not yet experienced core collapse, and they are in good
agreement with those quoted by \citet[][$r_c=26\farcs 3$ and
$c=1.71$]{mcL05}, and marginally consistent with those listed by
\citet[][$r_c=25\farcs 2$ and $c=1.83$]{har96}, both derived from the surface
brightness profile.  Our value of $r_c$ corresponds to $\sim 1$ pc assuming
the distance modulus $(m-M)_0=14.37$ \citep[$d\sim 7.5$ Kpc,][]{fe99b}.

\section{DEFINITION OF THE SAMPLES}
\label{sec:samples}
In order to study the BSS radial distribution and detect possible
peculiarities, both the BSS and a reference population must be properly
defined.  Since the HST and the WFI data sets have been observed in different
photometric bands, different selection boxes are needed to separate the
samples in the CMDs. The adopted strategy is described in the following
sections (see also Ferraro et al. 2004 for a detailed discussion of this
issue).

\subsection{The BSS selection}
\begin{center}
\begin{figure}[!p]
\includegraphics[scale=0.7]{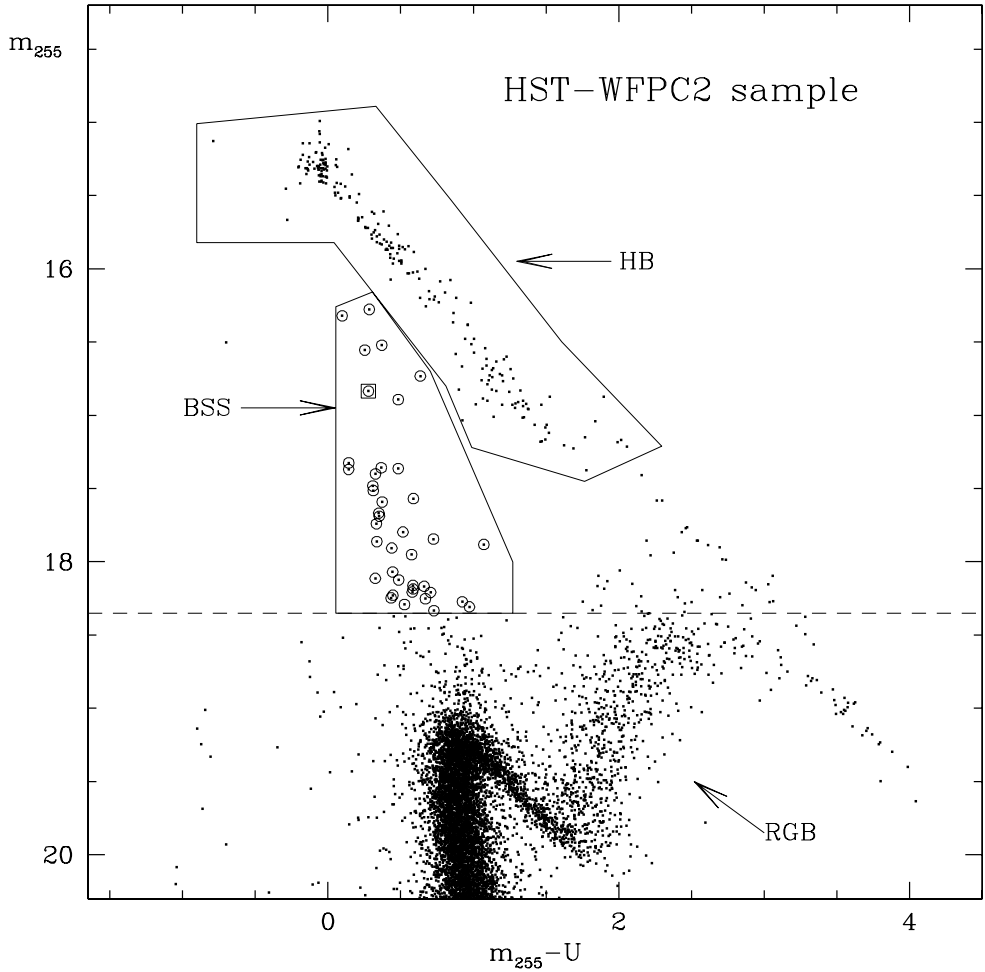}
\caption{CMD of the ultraviolet HST sample.  The adopted magnitude limit and
selection box used for the definition of the BSS population are shown.  The
resulting fiducial BSS are marked with empty circles. The open square
corresponds to the variable BSS identified by \citet{dris98}. The box adopted
for the selection of HB stars is also shown.}
\label{fig:UVsel}
\end{figure}
\end{center}

\begin{center}
\begin{figure}[!p]
\includegraphics[scale=0.7]{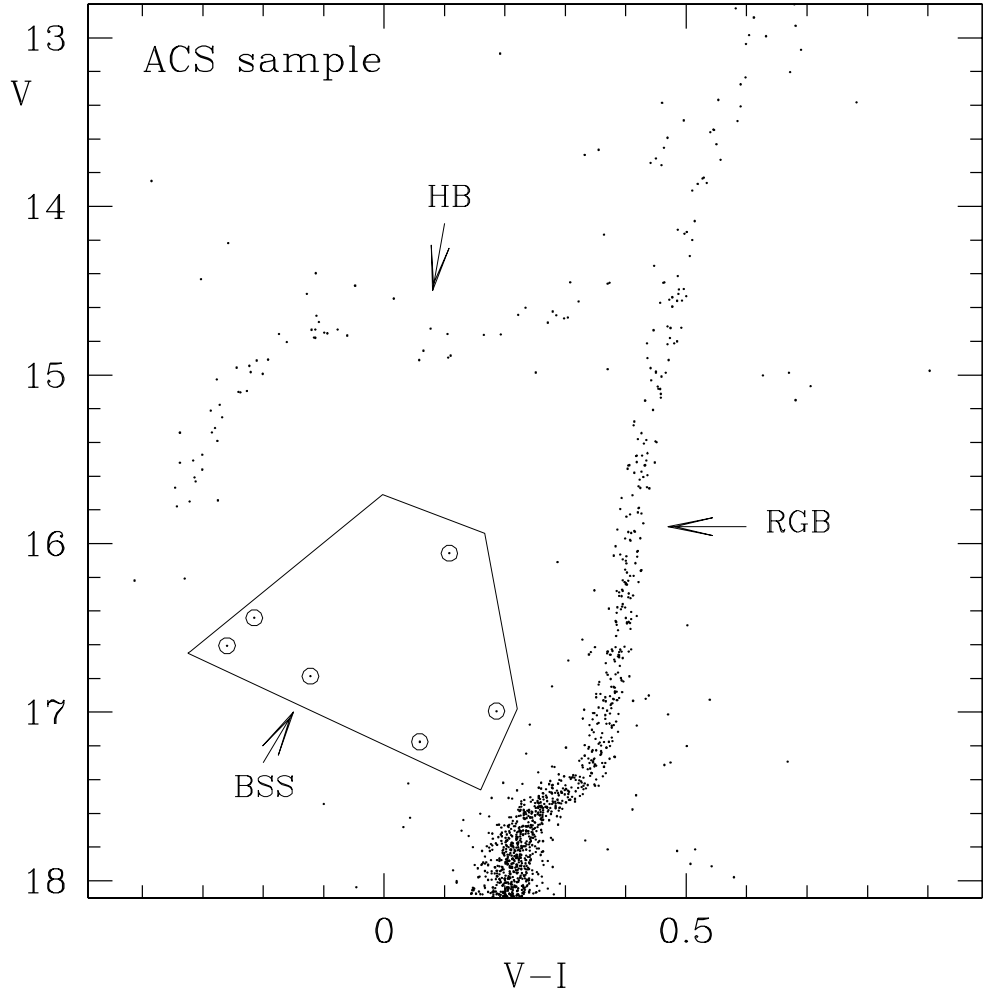}
\caption{CMD of the ACS complementary sample.  The BSS selection box is
shown, and the resulting fiducial BSS are marked with empty circles.}
\label{fig:ACSsel}
\end{figure}
\end{center}

At UV wavelengths BSS are among the brightest objects in a GC, and RGB stars
are particularly faint.  By combining these advantages with the
high-resolution capability of HST, the usual problems
associated with photometric blends and crowding in the high density central
regions of GCs are minimized, and BSS can be most reliably recognized and
separated from the other populations in the UV CMDs.
For these reasons our primary criterion for the definition of the BSS sample
is based on the position of stars in the ($m_{255},~m_{255}-U$) plane.  In
order to avoid incompleteness bias and the possible contamination from TO and
sub-giant branch stars, we have adopted a limiting magnitude $m_{255}=18.35$,
roughly corresponding to 1 magnitude brighter than the cluster TO. This is
also the limiting magnitude used by W06, facilitating the comparison with
their study. The resulting BSS selection box in the UV CMD is shown in Figure
\ref{fig:UVsel}.
Once selected in the UV CMD, the bulk of the BSS lying in the field in common
with the optical-HST sample has been used to define the selection box and the
limiting magnitude in the ($B,~B-V$) plane.  The latter turns out to be
$B\simeq 17.85$, and the adopted BSS selection box in the optical CMD is
shown in Figure \ref{fig:Optsel}.  The two stars lying outside the selection
box (namely BSS-19 and BSS-20 in Table \ref{tab:BSS}) have been identified
as BSS from the ($m_{255},~m_{255}-U$) CMD.  Indeed, they are typical
examples of how the optical magnitudes are prone to blend/crowding problems,
while the BSS selection in UV bands is much more secure and reliable.  An
additional BSS (BSS-47 in Table \ref{tab:BSS}) lies near the edge of the ACS
FoV and has only $V$ and $I$ observations; thus it was selected in the
$(V,~V-I)$ plane (see Figure \ref{fig:ACSsel}, where this BSS is shown
together with the other 5 identified in the ACS complementary sample).

With these criteria we have identified 60 BSS: 47 BSS in the HST sample
($r\le 115\arcsec$) and 13 in the WFI one. Their coordinates and magnitudes
are listed in Table \ref{tab:BSS}.  Out of the 47 BSS identified in the HST
sample, 41 are from the WFPC2 data set, and 6 from the ACS catalog. As shown
in Figure \ref{fig:HST} their projected distribution is quite asymmetric with
the N-E sector seemingly underpopulated. The statistical significance of such
an asymmetry appears even higher if only the BSS outside the core are
considered.  However a quantitative discussion of this topic is not warranted
unless additional evidences supporting this anomalous spatial distribution
are collected.  One of the inner BSS (BSS-29 in Table \ref{tab:BSS}) lying at
$21\farcs 76$ from the center, corresponds to the low-amplitude variable
HST-V28 identified by \citet{dris98}\footnote{The observations presented here
do not have the time coverage needed to properly search for BSS
variability.}.  In the WFI sample ($r >115\arcsec$) we find 13 BSS, with a
more symmetric spatial distribution (see Figure \ref{fig:WFI}).  The most
distant BSS (BSS-60 in Table \ref{tab:BSS}, marked with an empty triangle in
Fig.\ref{fig:Optsel}) lies at $\sim 24\arcmin$ from the center, i.e., beyond
the cluster tidal radius. Hence, it might be an evaporating BSS previously
belonging to the cluster. However, further investigations are needed before
firmly assessing this issue.

In order to perform a proper comparison with W06 study, we have transformed
their BSS catalog in our astrometric system, and we have found that 50 BSS of
their bright sample lie at $r\le 115\arcsec$: 35 are from the HST sample, 13
from the Canada France Hawaii Telescope (CFHT) data set, and 2 from the Cerro
Tololo Inter-American Observatory (CTIO) sample; in the outer regions
($115\arcsec < r \lsim 425\arcsec$) 9 BSS are identified, all from the CTIO
data set.

\begin{figure}[!p]
\begin{center}
\includegraphics[scale=1.]{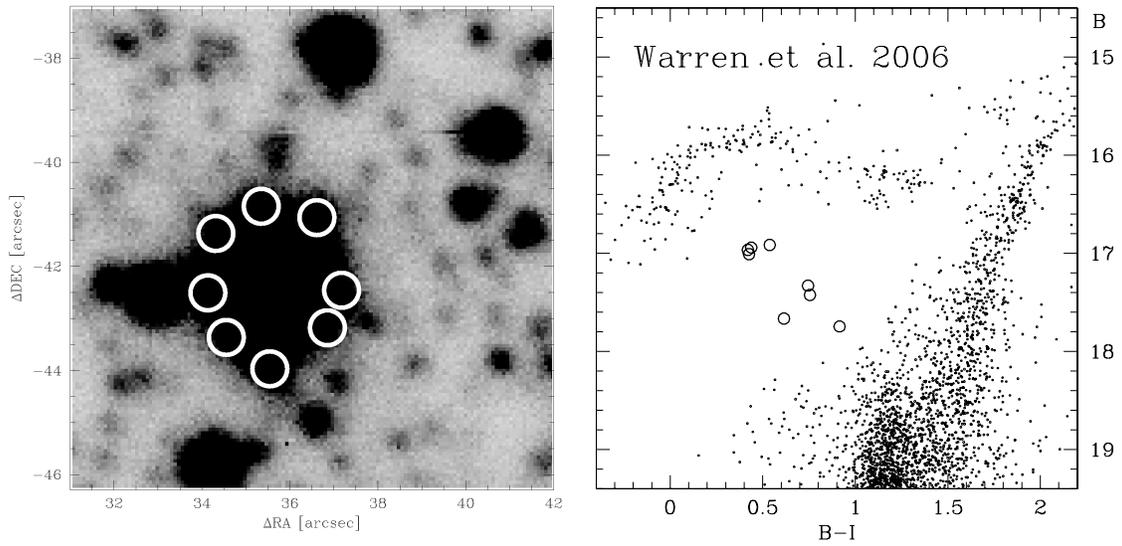}
\caption{{\it Left-hand panel}: position of the 8 false BSS (marked with
white circles) as derived from Table 2 of W06, overplotted to the CFHT image
(units are the same as in their Figure 1). As can be seen, a heavily
saturated star is responsible for the false identification.  {\it Right-hand
panel}: location of the 8 false BSS (empty circles) in the $(B,\, B-I)$
plane, as derived from Table 2 of W06 (cfr. to their Fig. 2).}
\label{fig:satura}
\end{center}
\end{figure}

By cross correlating W06 bright sample with our catalog we have found 43 BSS
in common (see Table \ref{tab:BSS}), 37 at $r\le 115\arcsec$ and 6 outward.
In particular, 33 BSS out of the 41 (i.e., 80\% of the total) that we have
identified in the WFPC2-HST sample\footnote{Note that the WFPC2-HST
observations used in W06 and in the present study are the same.} are found in
both catalogs, while 3 of our BSS belong 5 to their faint BSS sample (namely,
BSS-27, 34, and 40, corresponding to their Core BSS 70, 79, and 76,
respectively), 5 of our BSS have been missed in W06 paper, and 2 objects in
their sample are classified as HB stars in our study.  This is probably due
to different selection criteria, and/or small differences in the measured
magnitudes, caused by the different data reduction procedures and photometric
analysis. For example, W06 identify the BSS on the basis of both the UV and
the optical observations, while we select the BSS only in the UV plane
whenever possible.  Out of the other 15 BSS found at $r\le115\arcsec$ in the
ground-based CFHT/CTIO sample of W06, 8 BSS (Core BSS 38--45 in their Table
2) clearly are false identifications. They are arranged in a very unlikely
ring around a strongly saturated star, as can be seen in Figure
\ref{fig:satura}, where the position in the sky of the 8 spurious BSS are
overplotted on the CFHT image.  Though they clearly are spurious
identifications, they still define a clean sequence in the $(B,~B-I)$ CMD,
nicely mimicking the BSS magnitudes and colors.  As already discussed in
previous papers, this once again demonstrates how automatic procedures for
the search of peculiar objects are prone to errors, especially when using
ground-based observations to probe very crowded stellar regions.  We
emphasize that all the candidate BSS listed in our Table \ref{tab:BSS} have
been visually inspected evaluating the quality and the precision of the PSF
fitting. This procedure significantly reduces the possibility of introducing
spurious objects in the sample.  Out of the remaining 7 BSS, 4 objects
(namely their Core BSS 32, 30, 37 and 28) are also confirmed by our ACS
observations (BSS-42, 43, 44, and 45 respectively), while 2 others (their
Core BSS 27 and Ground BSS 6) are not found in the ACS data set, and the
remaining one (their Ground BSS 7) is not included in our observation FoV. In
turn, two BSS identified in our ACS data set (BSS 46 and 47) are missed in
their sample.  Concerning the BSS lying at $115\arcsec < r < 450\arcsec$, 6
objects (out of 9 found in both samples) are in common between the two
catalogs (see Table \ref{tab:BSS}), one (BSS-55) belongs to W06 faint sample
(their Ground BSS 23), while the remaining 2 do not coincide. Moreover, 4
additional BSS have been identified at $r > 450\arcsec$ in our study.

\subsection{The reference population}
\label{sec:HB}
Since the HB sequence is bright and well separable in the UV and optical
CMDs, we chose these stars as the primary representative population of normal
cluster stars to be used for the comparison with the BSS data set.  As with
the BSS, the HB sample was first defined in the ($m_{255},~m_{255}-U$) plane,
and the corresponding selection box in ($B,~B-V$) has then been determined by
using the stars in common between the UV and the optical samples. The
resulting selection boxes in both diagrams are shown in Figures
\ref{fig:UVsel} and \ref{fig:Optsel}, and are designed to include the bulk of
HB stars\footnote{The large dispersion in the redder HB stars arises because
RR Lyrae variables are included.}. Slightly different selection boxes would
include or exclude a few stars only without affecting the results.

We have used WFI observations to roughly estimate the impact of possible
foreground field stars contamination on the cluster population selection. As
shown in the right-hand panel of Figure \ref{fig:Optsel}, field stars appear
to define an almost vertical sequence at $0.4 < B-V < 1$ in the ($B,~B-V$)
CMD. Hence, they do not affect the BSS selection box, but marginally
contaminate the reddest end of the HB. In particular, 5 objects have been
found to lie within the adopted HB box in the region at $r>r_t$ sample by our
observations ($\sim 194$ arcmin$^2$); this corresponds to $0.026$ spurious HB
stars per arcmin$^2$. On the basis of this, 11 field stars are expected to
"contaminate" the HB population over the sampled cluster region ($r<r_t$).

\section{THE BSS RADIAL DISTRIBUTION}
\label{sec:radist}
The radial distribution of BSS in M5 has been studied following the
same procedure previously adopted for other clusters (see references
in Ferraro 2006; Beccari et al. 2006).

\begin{figure}[!p]
\begin{center}
\includegraphics[scale=0.7]{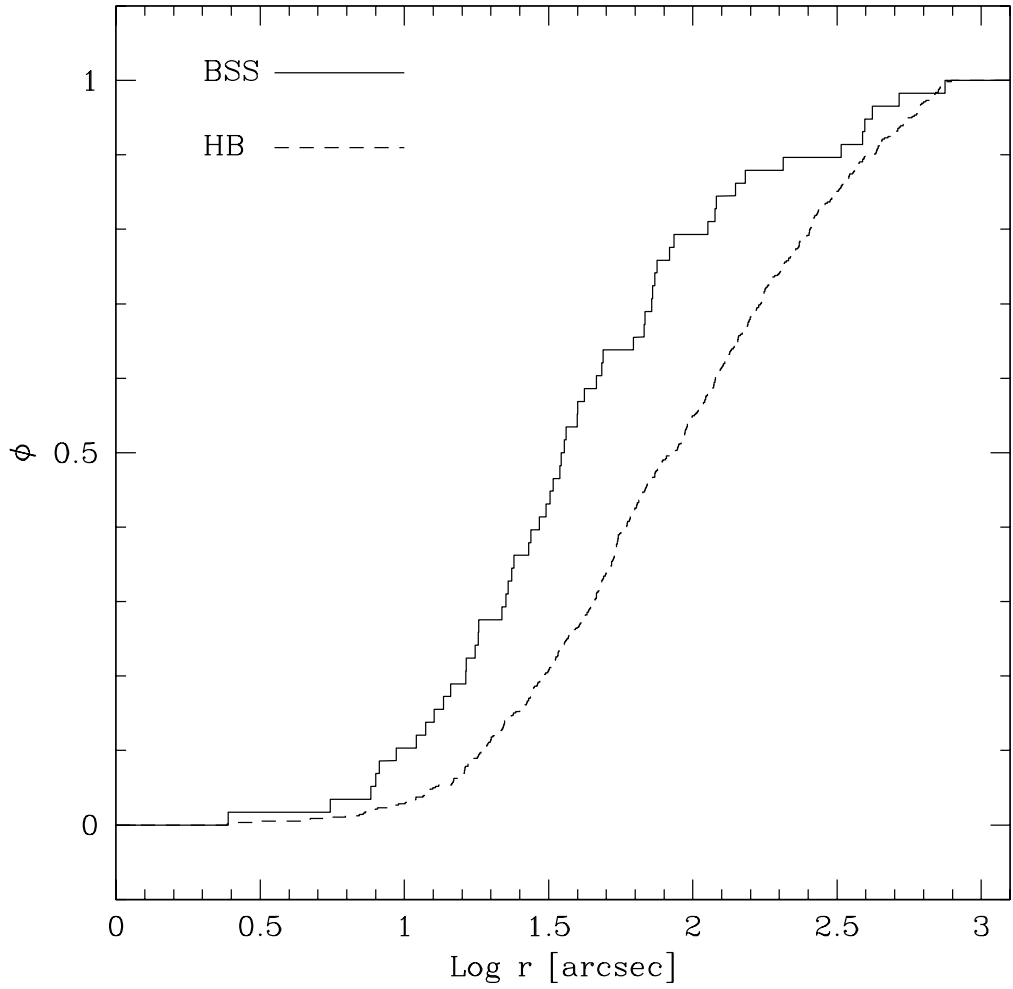}
\caption{Cumulative radial distribution of BSS (solid line) and HB stars
(dashed line) as a function of the projected distance from the cluster center
for the combined HST+WFI sample. The two distributions differ at $\sim 4
\sigma$ level.}
\label{fig:KS}
\end{center}
\end{figure}

First, we have compared the BSS cumulative radial distribution to that of HB
stars. A Kolmogorov-Smirnov test gives a $\sim 10^{-4}$ probability that they
are extracted from the same population (see Figure \ref{fig:KS}). BSS are
more centrally concentrated than HB stars at $\sim 4\sigma$ level.

For a more quantitative analysis, the surveyed area has been divided into 8
concentric annuli, with radii listed in Table \ref{tab:annuli}.  The number
of BSS ($N_{\rm BSS}$) and HB stars ($N_{\rm HB}$), as well as the fraction
of sampled luminosity ($L^{\rm samp}$) have been measured in the 8 annuli and
the obtained values are listed in Table \ref{tab:annuli}.  Note that HB star
counts listed in the table are already decontaminated from field stars,
according to the procedure described in Section \ref{sec:HB} (1, 2, and 8 HB
stars in the three outer annuli have been estimated to be field stars).  The
listed values have been used to compute the specific frequency $F_{\rm
BSS}^{\rm HB}\equiv N_{\rm BSS}/N_{\rm HB}$, and the double normalized ratio
(see Ferraro et al. 1993):
\begin{equation}
R_{\rm pop}=\frac{(N_{\rm pop}/N_{\rm pop}^{\rm tot})}{(L^{\rm samp}/L_{\rm
tot}^{\rm samp})},
\label{eq:spec_freq}
\end{equation}
with ${\rm pop} =$ BSS, HB.

In the present study luminosities have been calculated from the surface
density profile shown in Figure \ref{fig:prof}.  The surface density has been
transformed into luminosity by means of a normalization factor obtained by
assuming that the value obtained in the core ($r\le 27\arcsec$) is equal to
the sum of the luminosities of all the stars with $V\le 20$ lying in this
region. The distance modulus quoted in Section \ref{sec:dens_prof} and a
reddening $E(B-V)=0.03$ have been adopted \citep[][]{fe99b}. The fraction of
area sampled by the observations in each annulus has been carefully computed,
and the sampled luminosity in each annulus has been corrected for incomplete
spatial coverage (in the case of annuli 3 and 8; see Figures \ref{fig:HST}
and \ref{fig:WFI}).

The resulting radial trend of $R_{\rm HB}$ is essentially constant with a
value close to unity over the surveyed area (see Figure \ref{fig:Rpop}). This
is just what expected on the basis of the stellar evolution theory, which
predicts that the fraction of stars in any post-main sequence evolutionary
stage is strictly proportional to the fraction of the sampled luminosity
(Renzini \& Fusi Pecci 1988).  Conversely, BSS follow a completely different
radial distribution.  As shown in Figure \ref{fig:Rpop} the specific
frequency $R_{\rm BSS}$ is highly peaked at the cluster center (a factor of
$\sim 3$ higher than $R_{\rm HB}$ in the innermost bin), decreases to a
minimum\footnote{Note that no BSS have been found between $3.\arcmin 5$ and
$5\arcmin$.} at $r\simeq 10\, r_c$, and rises again outward.  The same
behavior is clearly visible also in Figure \ref{fig:simu}, where the
population ratio $N_{\rm BSS}/N_{\rm HB}$ is plotted as a function of
$r/r_c$.

Note that the region between $800\arcsec$ and $r_t\simeq 1290\arcsec$ (and
thus also BSS-59, that lies at $r\simeq 995\farcs 5$) has not been considered
in the analysis, since our observations provide a poor sampling of this
annulus: only 35\% of its area, corresponding to $\sim 0.4\%$ of the total
sampled light, is covered by the WFI pointing. However, for sake of
completeness, we have plotted in Figure \ref{fig:cfr_radist} the
corresponding value of $F_{\rm BSS}^{\rm HB}$ even for this annulus (empty
circle in the upper panel): as can be seen, there is a hint for a flattening
of the BSS radial distribution in the cluster outskirts.

\subsection{Dynamical simulations}
\label{sec:simu}
Following the same approach as \citet[][2006]{ma04}, we now exploit dynamical
simulations to derive some clues about the BSS formation mechanisms from
their observed radial distribution. We use the Monte-Carlo simulation code
originally developed by Sigurdsson \& Phinney (1995) and upgraded in
\citet[][2006]{ma04}.  In any simulation run we follow the dynamical
evolution of $N$ BSS within a background cluster, taking into account the
effects of both dynamical friction and distant encounters.  We identify as
COL-BSS those objects having initial positions $r_i \lsim r_c$, and as MT-BSS
stars initially lying at $r_i \gg r_c$ (this because stellar collisions are
most probable in the central high-density regions of the cluster, while
primordial binaries most likely evolve in isolation in the periphery).
Within these two radial ranges, all initial positions are randomly generated
following the probability distribution appropriate for a King model.  The BSS
initial velocities are randomly extracted from the cluster velocity
distribution illustrated in Sigurdsson \& Phinney (1995), and an additional
natal kick is assigned to the COL-BSS in order to account for the recoil
induced by the encounters.  Each BSS has characteristic mass $M$ and lifetime
$t_{\rm last}$. We follow their dynamical evolution in the cluster (fixed)
gravitational potential for a time $t_i$ ($i=1,N$), where each $t_i$ is a
randomly chosen fraction of $t_{\rm last}$.  At the end of the simulation we
register the final positions of BSS, and we compare their radial distribution
with the observed one. We repeat the procedure until a reasonable agreement
between the simulated and the observed distributions is reached; then, we
infer the percentage of collisional and mass-transfer BSS from the
distribution of the adopted initial positions in the simulation.

For a detailed discussion of the ranges of values appropriate for these
quantities and their effects on the final results we refer to \citet{ma06}.
Here we only list the assumptions made in the present
study:
\begin{itemize}
\item[--]the background cluster is approximated with a multi-mass King model,
determined as the best fit to the observed profile\footnote{By adopting the
same mass groups as those of \citet{ma06}, the resulting value of the King
dimensionless central potential is $W_0=9.7$}. The cluster central velocity
dispersion is set to $\sigma= 6.5\,{\rm km\,s^{-1}}$ (Dubath et al. 1997),
and, assuming $0.5 M_\odot$ as the average mass of the cluster stars, the
central stellar density is $n_c=2\times 10^4$ pc$^{-3}$ (Pryor \& Meylan
1993);
\item[--]the COL-BSS are distributed with initial positions $r_i\le r_c$ and
are given a natal kick velocity of $1\times\sigma$;
\item[--]initial positions ranging between $5\,r_c$ and $r_t$ (with the tidal
radius $r_t\simeq 48\, r_c$) have been considered for MT-BSS in different
runs; 
\item[--]BSS masses have been fixed to $M=1.2\, M_\odot$ \citep{fe06a}, and
their characteristic lifetime to $t_{last}=2$ Gyr;
\item[--]in each simulation run we have followed the evolution of $N=10,000$
BSS.
\end{itemize}

The simulated radial distribution that best reproduces the observed one (with
a reduced $\chi^2\simeq 0.6$) is shown in Figure~\ref{fig:simu} (solid line)
and is obtained by assuming that $\sim 80\%$ of the BSS population was formed
in the core through stellar collisions, while only $\sim 20\%$ is made of
MT-BSS.  A higher fraction ($\gsim 40\%$) of MT-BSS does not correctly
reproduce the steep decrease of the distribution and seriously overpredict
the number of BSS at $r\sim 10\, r_c$, where no BSS at all are found, but it
nicely matches the observed upturning point at $r\simeq 13\,r_c$ (see the
dashed line in Figure~\ref{fig:simu}). On the other hand, a population of
only COL-BSS is unable to properly reproduce the external upturn of the
distribution (see the dotted line in Figure~\ref{fig:simu}), and 100\% of
MT-BSS is also totally excluded.  Assuming heavier BSS (up to $M=1.5\,
M_\odot$) or different lifetimes $t_{last}$ (between 1 and 4 Gyr) does not
significantly change these conclusions, since both these parameters mainly
affect the external part of the simulated BSS distribution. Thus, an
appreciable effect can be seen only in the case of a relevant upturn, and
negligible variations are found in the best-fit case and when assuming 100\%
COL-BSS. The effect starts to be relevant in the simulations with 40\% or
more MT-BSS, which are however inconsistent with the observations at
intermediate radii (see above).

By using the simulations and the dynamical friction timescale \citep[from,
e.g.,][]{ma06}, we have also computed the radius of avoidance of M5. This is
defined as the characteristic radial distance within which all MT-BSS are
expected to have already sunk to the cluster core, because of mass
segregation processes. Assuming 12 Gyr for the age of M5 (Sandquist et
al. 1996) and $1.2\,M_\odot$ for the BSS mass, we find that $r_{\rm
avoid}\simeq 10\,r_c$.  This nicely corresponds to the position of the
minimum in the observed BSS radial distribution, in agreement with the
findings of Mapelli et al. (2004, 2006).

\section{SUMMARY AND DISCUSSION}
In this paper we have used a combination of HST UV and optical images of the
cluster center and wide-field ground-based observations covering the entire
cluster extension to derive the main structural parameters and to study the
BSS population of the galactic globular cluster M5.

The accurate determination of the cluster center of gravity from the
high-resolution data gives $\alpha({\rm J2000}) = 15^{\rm h}\, 18^{\rm m}\,
33\fs 53$, $\delta({\rm J2000}) = +2^{\rm o}\, 4\arcmin\, 57\farcs 06$, with
a 1$\sigma$ uncertainty of $0\farcs 5$ in both $\alpha$ and
$\delta$.  The cluster density profile, determined from
direct star counts, is well fit by a King model with core radius
$r_c=27\arcsec$ and concentration $c=1.68$, thus suggesting that M5 has not
yet suffered the core collapse.

\begin{figure}[!hp]
\begin{center}
\includegraphics[scale=0.7]{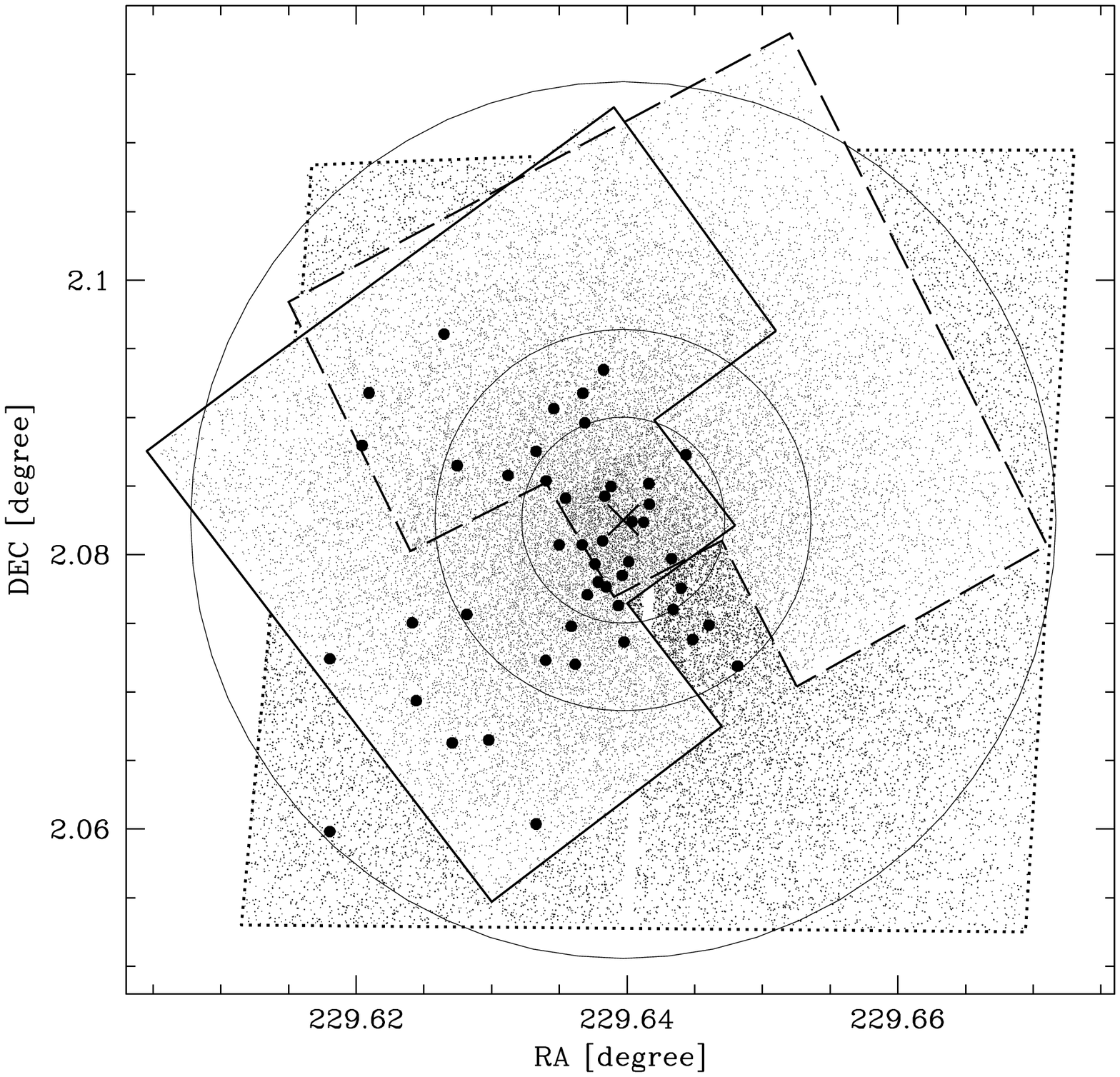}
\caption{Map of the HST sample. The heavy solid line delimits the HST-WFPC2
FoV of our UV observations (Prop. 6607), the dashed line bounds the FoV of
the optical HST-WFPC2 observations by Piotto et al. (2002), and the dotted
line marks the edge of the complementary ACS data set. The derived center of
gravity $C_{\rm grav}$ is marked with a cross.  BSS (heavy dots) and the
concentric annuli used to study their radial distribution (cfr. Table
\ref{tab:annuli}) are also shown. The inner and outer annuli correspond to
$r=r_c=27\arcsec$ and $r=115\arcsec$, respectively.}
\label{fig:HST}
\end{center}
\end{figure}

\begin{figure}[!p]
\begin{center}
\includegraphics[scale=0.7]{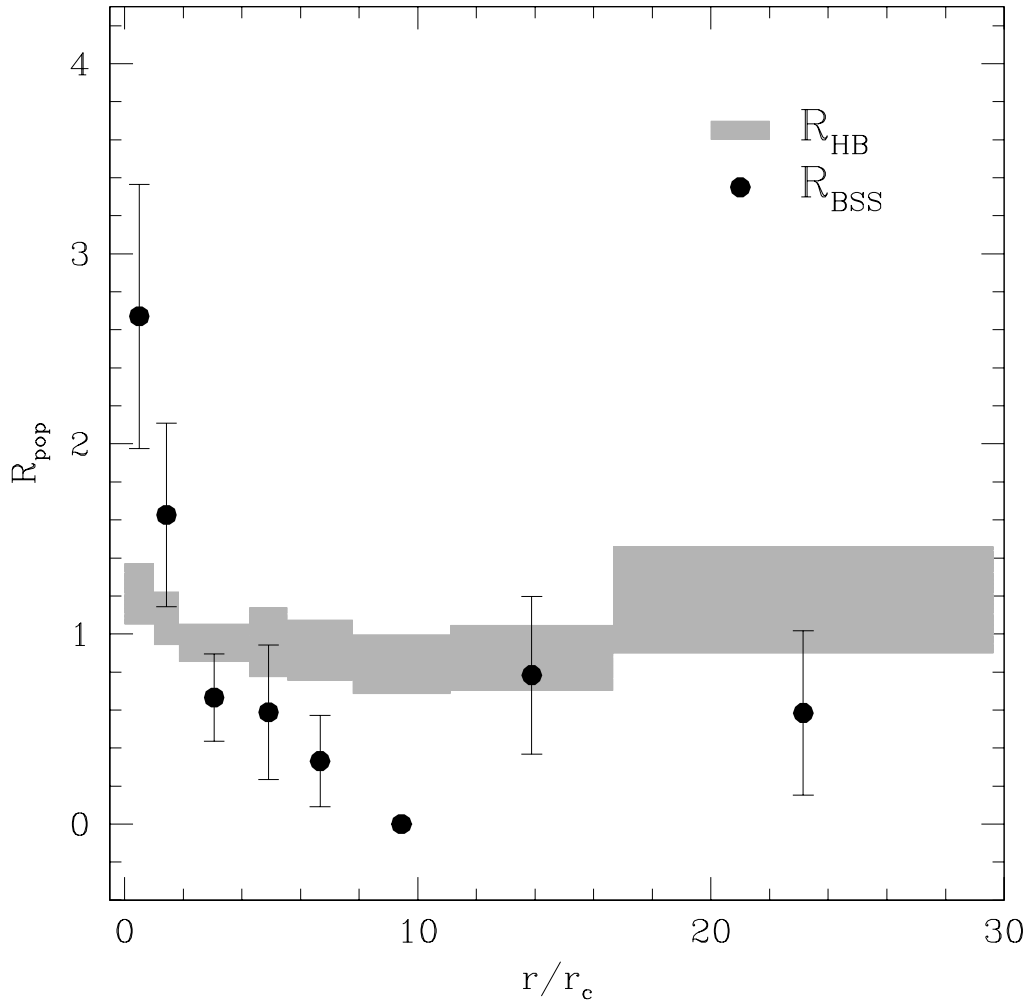}
\caption{Radial distribution of the BSS and HB double normalized ratios, as
defined in equation~(\ref{eq:spec_freq}), plotted as a function of the radial
coordinate expressed in units of the core radius. $R_{\rm HB}$ (with the size
of the rectangles corresponding to the error bars computed as described in
Sabbi et al. 2004) is almost constant around unity over the entire cluster
extension, as expected for any normal, non-segregated cluster
population. Instead, the radial trend of $R_{\rm BSS}$ (dots with error bars)
is completely different: highly peaked in the center (a factor of $\sim$ 3
higher than $R_{\rm HB}$), decreasing at intermediate radii, and rising again
outward.}
\label{fig:Rpop}
\end{center}
\end{figure}

\begin{figure}[!p]
\begin{center}
\includegraphics[scale=0.8]{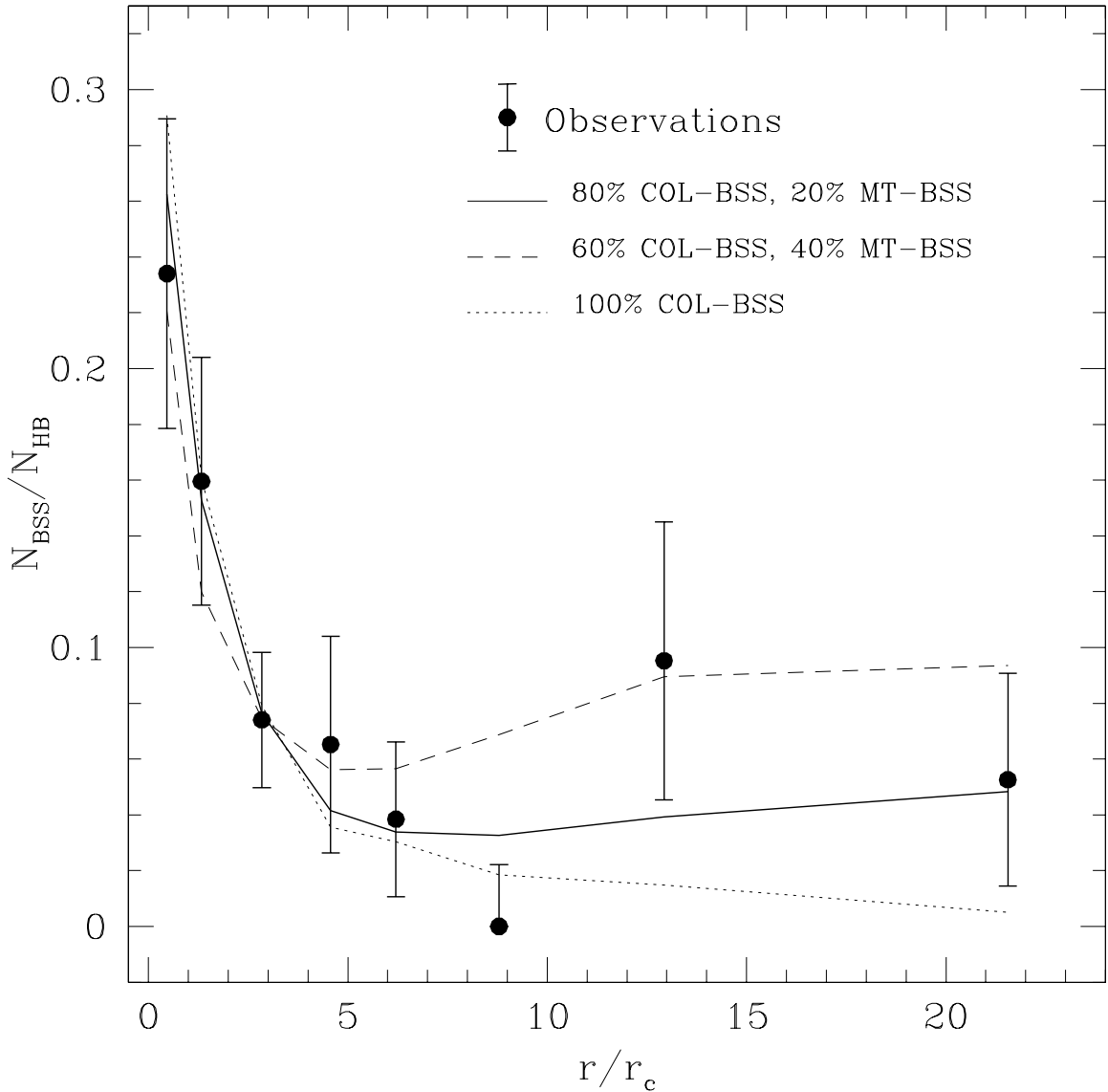}
\caption{Observed radial distribution of the specific frequency $N_{\rm
BSS}/N_{\rm HB}$ (filled circles with error bars), as a function of $r/r_c$.
The simulated distribution that best reproduces the observed one is shown as
a solid line and is obtained by assuming $80\%$ of COL-BSS and $20\%$ of
MT-BSS.  The simulated distributions obtained by assuming $40\%$ of MT-BSS
(dashed line) and $100\%$ COL-BSS (dotted line) are also shown.
}
\label{fig:simu}
\end{center}
\end{figure}

The BSS population of M5 amounts to a total of 59 objects, with a quite
asymmetric projected distribution (see Figure \ref{fig:HST}) and a high
degree of segregation in the cluster center.  With respect to the sampled
luminosity and to HB stars, the BSS radial distribution is bimodal: highly
peaked at $r\lsim r_c$, decreasing to a minimum at $r\simeq 10\, r_c$, and
rising again outward (see Figures \ref{fig:Rpop} and \ref{fig:simu}).

The comparison with results of W06 has revealed that 43 (out of 59) bright
BSS identified by these authors at $r\lsim 450\arcsec$ are in common with our
sample.  Moreover, 4 additional stars classified as faint BSS in their study
are in common with our BSS sample at $r\lsim 450\arcsec$. Considering that we
find 56 BSS within the same radial distance from the center, this corresponds
to 84\% matching of our catalogue.  The discrepancies are explained by
different data reduction procedures, photometric analysis, and adopted
selection criteria, other than the spurious identification of 8 BSS by W06,
due a strongly saturated star in their sample.  The central peak of the
$R_{\rm BSS}$ distribution in our study is slightly higher (but compatible
within the error bar) compared to that of W06, and we extend the analysis to
larger distance from the center (out to $r > 800\arcsec$), thus unveiling the
external upturn and the possible flattening of the BSS distribution in the
cluster outskirts.

\begin{figure}[!p]
\begin{center}
\includegraphics[scale=0.8]{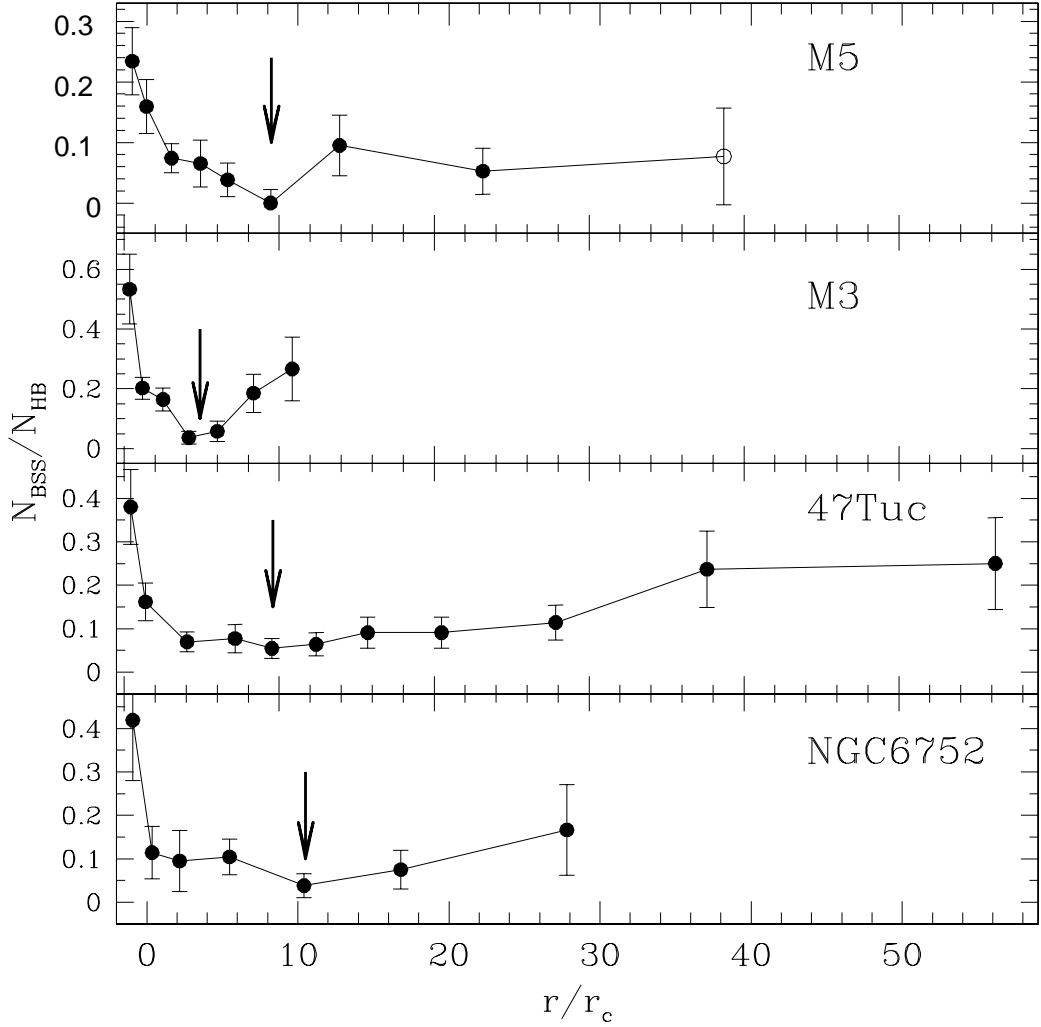}
\caption{Radial distribution of the population ratio $N_{\rm BSS}/N_{\rm HB}$
for M5, M3, 47~Tuc, and NGC~6752, plotted as a function of the radial
distance from the cluster center, normalized to the core radius $r_c$ (from
Mapelli et al. 2006, $r_c\simeq 30\arcsec, 21\arcsec, 28\arcsec$ for M3,
47~Tuc, and NGC~6752, respectively). The arrows indicate the position of the
minimum of the distribution in each case. The outermost point shown for M5
(empty circle) corresponds to BSS-58, lying at $r\simeq 995\arcsec$. This
star has not been considered in the quantitative study of the BSS radial
distribution since only a negligible fraction of the annuls between
$800\arcsec$ and $r_t$ is sampled by our observations.}
\label{fig:cfr_radist}
\end{center}
\end{figure}

Moreover, we have compared the BSS radial distribution of M5 with that
observed in other GCs studied in a similar way. In Figure
\ref{fig:cfr_radist} we plot the specific frequency $F_{\rm BSS}^{\rm HB}$ as
a function of $(r/r_c)$ for M5, M3, 47~Tuc, and NGC~6752.  Such a comparison
shows that the BSS radial distributions in these clusters are only
\emph{qualitatively} similar, with a high concentration at the center and an
upturn outward. However, significant quantitative differences are apparent:
(1) the $F_{\rm BSS}^{\rm HB}$ peak value, (2) the steepness of the
decreasing branch of the distribution, (3) the radial position of the minimum
(marked by arrows in the figure), and (4) the extension of the ``zone of
avoidance,'' i.e., the intermediate region poorly populated by BSS.  In
particular M5 shows the smallest $F_{\rm BSS}^{\rm HB}$ peak value: it turns
out to be $\sim 0.24$, versus a typical value $\gsim 0.4$ in all the other
cases. It also shows the mildest decreasing slope: at $r\approx 2\,r_c$ the
specific frequency in M5 is about a half of the peak value, while it
decreases by a factor of 4 in all the other clusters. Conversely, it is
interesting to note that the value reached by $F_{\rm BSS}^{\rm HB}$ in the
external regions is $\sim$ 50-60\% of the central peak in all the studied
clusters.  Another difference between M5 and the other systems concerns the
ratio between the radius of avoidance and the tidal radius: $r_{\rm
avoid}\simeq 0.2\,r_t$ for M5, while $r_{\rm avoid}\lsim 0.13\,r_t$ for
47~Tuc, M3, and NGC~6752 (see Tables 1 and 2 in Mapelli et al. 2006).

The dynamical simulations discussed in Section \ref{sec:simu} suggest that
the majority of BSS in M5 are collisional, with a content of MT-BSS ranging
between 20\% and 40\% of the overall population. This fraction seems to be
smaller than that (40-50\%) derived for M3, 47~Tuc and NGC~6752 by
\citet{ma06}, in qualitative agreement with the smaller value of $r_{\rm
avoid}/r_t$ estimated for M5, which indicates that the fraction of cluster
currently depopulated of BSS is larger in this system than in the other
cases. More in general, the results shown in Figure \ref{fig:simu} exclude a
pure collisional BSS content for M5.

\begin{center}
\begin{figure}[!p]
\includegraphics[scale=0.7]{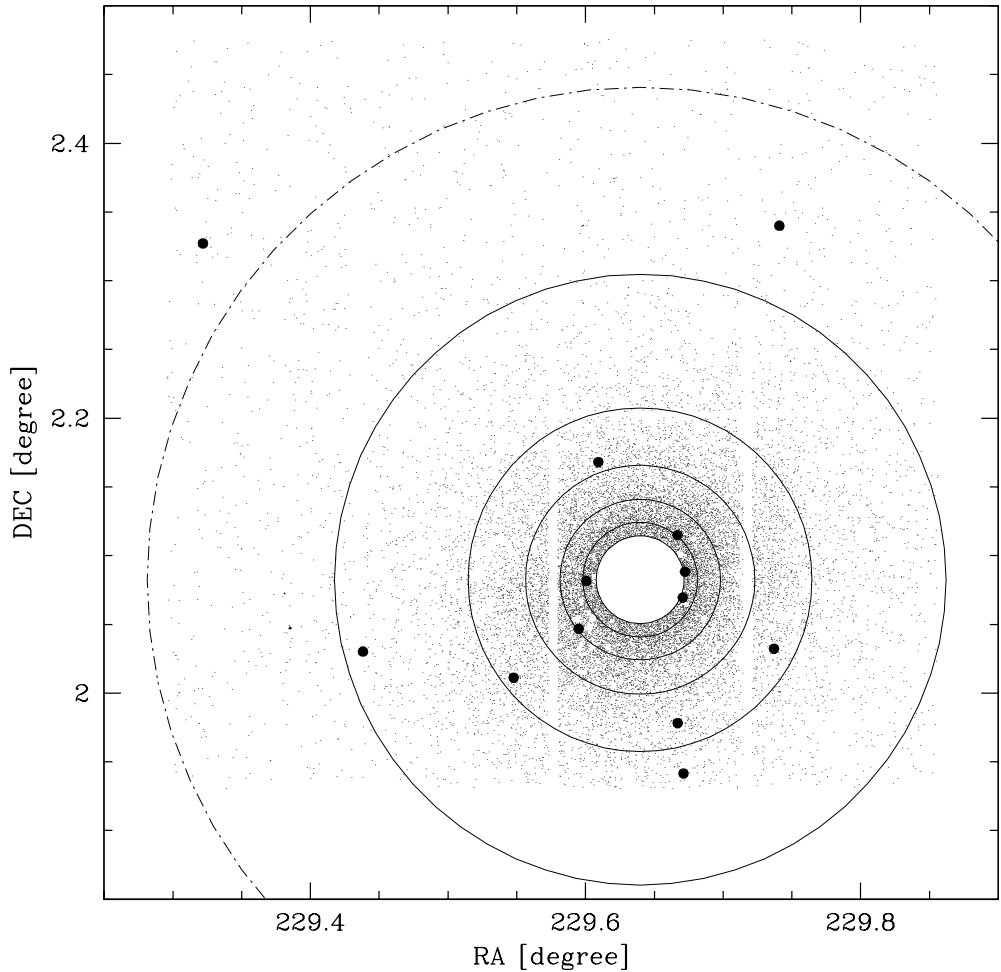}
\caption{Map of the WFI sample.  All BSS detected in the WFI sample are
marked as heavy dots, and the concentric annuli used to study their radial
distribution are shown as solid lines, with the inner and outer annuli
corresponding to $r=115\arcsec$ and $r=800\arcsec$, respectively (cfr. Table
\ref{tab:annuli}). The circle corresponding to the tidal radius ($r_t\simeq
21\farcm 5$) is also shown as dashed-dotted line. The BSS lying beyond $r_t$
might represent a BSS previously belonging to M5 and now evaporating from the
cluster.}
\label{fig:WFI}
\end{figure}
\end{center}

\begin{center}
\begin{figure}[!p]
\includegraphics[scale=0.7]{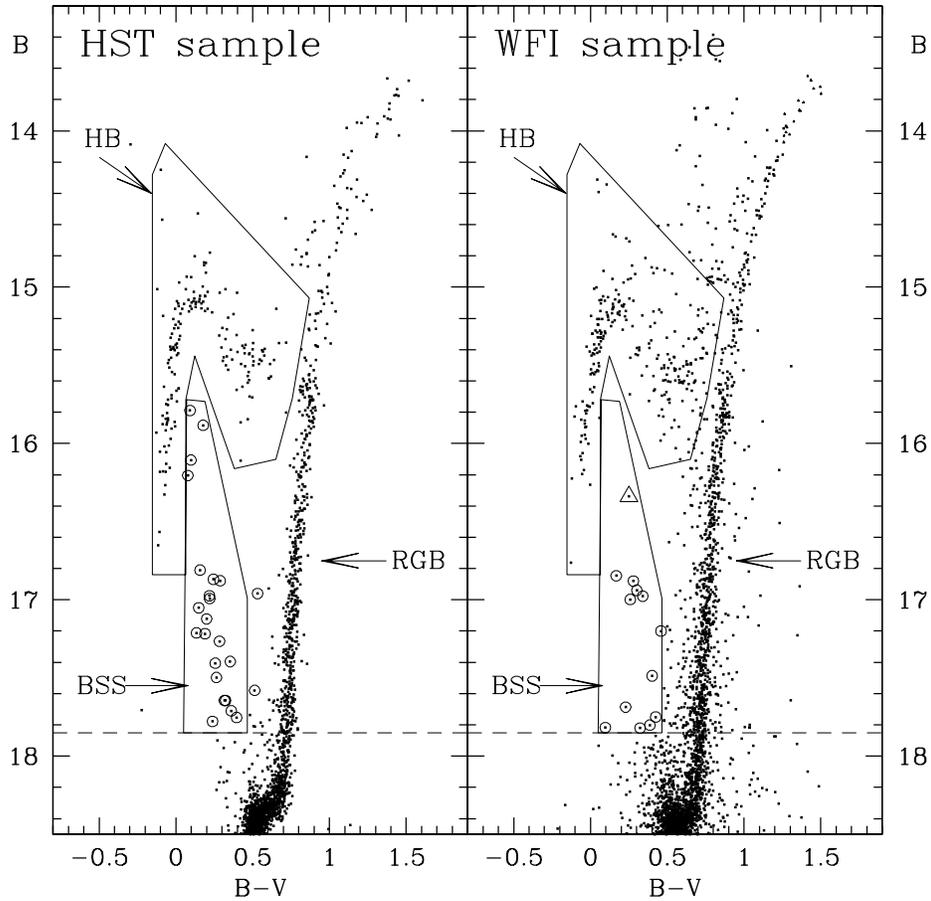}
\caption{CMD of the optical HST-WFPC2 and WFI samples.  The adopted BSS and
HB selection boxes are shown, and all the BSS identified in these samples are
marked with the empty circles. The two BSS not included in the box in the
left-hand panel lie well within the selection box in the UV plane and are
therefore considered as fiducial BSS. The empty triangle in the right-hand
panel corresponds to the BSS identified beyond the cluster tidal radius, at
$r\simeq 24\arcmin$.}
\label{fig:Optsel}
\end{figure}
\end{center}

Our study has also revealed the presence of a candidate BSS at $\sim
24\arcmin$ from the center, i.e., beyond the cluster tidal radius (see
Figures \ref{fig:WFI} and \ref{fig:Optsel} and BSS-59 in Table
\ref{tab:BSS}). If confirmed, this could represent a very interesting case of
a BSS previously belonging to M5 and then evaporating from the cluster (a BSS
kicked off from the core the because of dynamical interactions?).

\acknowledgements{This research was supported by Agenzia Spaziale Italiana
under contract ASI-INAF I/023/05/0, by the Istituto Nazionale di Astrofisica
under contract PRIN/INAF 2006, and by the Ministero dell'Istruzione,
dell'Universit\`a e della Ricerca. RTR is partially funded by NASA through
grant number GO-10524 from the Space Telescope Science Institute. We thank
the referee E. Sandquist for the careful reading of the manuscript and the
useful comments and suggestions that significantly improved the presentation
of the paper.}

\begin{deluxetable}{lcccccccl}
\footnotesize
\tablewidth{15cm}
\tablecaption{The BSS population of M5}
\startdata \\
\hline \hline
Name   &    RA       &   DEC     &$m_{255}$& U   &   B   &   V   &  I  & W06    \\
       & [degree]    &  [degree] &       &       &       &       &     &        \\
\hline
BSS-1  & 229.6354506 & 2.0841090 & 16.52 & 16.15 & 15.88 & 15.71 &   -   & CR2 \\ 
BSS-2  & 229.6388102 & 2.0849660 & 17.95 & 17.38 & 17.40 & 17.04 &   -   & CR4 \\ 
BSS-3  & 229.6383433 & 2.0842640 & 18.21 & 17.63 & 17.64 & 17.32 &   -   & CR3 \\ 
BSS-4  & 229.6416234 & 2.0851791 & 17.59 & 17.22 & 17.05 & 16.90 &   -   & CR5 \\ 
BSS-5  & 229.6416518 & 2.0836794 & 16.28 & 15.99 & 15.79 & 15.70 &   -   & CR1 \\ 
BSS-6  & 229.6381953 & 2.0810119 & 17.36 & 16.99 & 16.81 & 16.65 &   -   & CR21 \\ 
BSS-7  & 229.6403657 & 2.0824062 & 17.40 & 17.07 & 16.97 & 16.76 &   -   & CR12 \\ 
BSS-8  & 229.6412279 & 2.0823768 & 17.91 & 17.47 & 17.41 & 17.15 &   -   & CR13 \\ 
BSS-9  & 229.6376256 & 2.0793288 & 17.84 & 17.12 & 16.99 & 16.77 &   -   & CR23 \\ 
BSS-10 & 229.6401139 & 2.0794858 & 17.57 & 16.98 & 16.87 & 16.62 &   -   & CR22 \\ 
BSS-11 & 229.6396566 & 2.0784944 & 17.51 & 17.20 & 17.12 & 16.92 &   -   & CR24 \\ 
BSS-12 & 229.6432834 & 2.0797197 & 18.12 & 17.64 & 17.78 & 17.54 &   -   &  -   \\ 
BSS-13 & 229.6384406 & 2.0776614 & 17.36 & 16.88 & 16.88 & 16.59 &   -   & CR25 \\ 
BSS-14 & 229.6274500 & 2.0864896 & 18.07 & 17.63 & 17.64 & 17.33 &   -   & CR8 \\ 
BSS-15 & 229.6204246 & 2.0879629 & 18.33 & 17.61 & 17.75 & 17.36 &   -   & CR11 \\ 
BSS-16 & 229.6209379 & 2.0917858 & 17.80 & 17.28 & 17.26 & 16.98 &   -   & CR18 \\ 
BSS-17 & 229.6264834 & 2.0960870 & 16.32 & 16.22 & 16.20 & 16.13 &   -   & CR20 \\ 
BSS-18 & 229.6368731 & 2.0896002 & 16.56 & 16.30 & 16.11 & 16.01 &   -   & CR14 \\ 
BSS-19 & 229.6367309 & 2.0917639 & 18.27 & 17.35 & 17.58 & 17.07 &   -   & CR17 \\ 
BSS-20 & 229.6345837 & 2.0906438 & 17.88 & 16.81 & 16.96 & 16.43 &   -   & CR16 \\ 
BSS-21 & 229.6382677 & 2.0934706 & 18.25 & 17.58 & 17.71 & 17.35 &   -   & CR19 \\ 
BSS-22 & 229.6340227 & 2.0853879 & 17.67 & 17.32 & 17.22 & 17.03 &   -   & CR7 \\ 
BSS-23 & 229.6332685 & 2.0875294 & 17.69 & 17.34 & 17.21 & 17.08 &   -   & CR10 \\ 
BSS-24 & 229.6366685 & 2.0807168 & 18.23 & 17.78 & 17.67 & 17.37 &   -   &  -   \\ 
BSS-25 & 229.6393544 & 2.0762832 & 18.11 & 17.79 & 17.72 & 17.50 &   -   &  -   \\ 
BSS-26 & 229.6378381 & 2.0779999 & 17.86 & 17.52 & 17.43 & 17.27 &   -   &  -   \\ 
BSS-27 & 229.6349851 & 2.0807202 & 18.17 & 17.51 & 17.74 & 17.30 &   -   & CR70 \\ 
BSS-28 & 229.6397645 & 2.0736403 & 18.19 & 17.60 & 17.69 & 17.28 &   -   & CR33 \\ 
BSS-29 & 229.6370495 & 2.0770798 & 16.83 & 16.56 & 16.57 & 17.75 &   -   & CR26 \\ 
BSS-30 & 229.6358816 & 2.0747883 & 18.25 & 17.81 & 17.79 & 17.51 &   -   & CR31 \\ 
BSS-31 & 229.6361653 & 2.0720147 & 18.29 & 17.77 & 17.81 & 17.47 &   -   & CR36 \\ 
BSS-32 & 229.6339822 & 2.0723032 & 16.73 & 16.10 & 16.16 & 15.95 &   -   & CR35 \\ 
BSS-33 & 229.6281392 & 2.0756490 & 17.74 & 17.41 & 17.22 & 17.09 &   -   & CR29 \\ 
BSS-34 & 229.6241278 & 2.0750261 & 18.21 & 17.50 & 17.65 & 17.27 &   -   & CR79 \\ 
BSS-35 & 229.6332759 & 2.0603761 & 17.48 & 17.17 & 16.95 & 16.86 &   -   & CR48 \\ 
BSS-36 & 229.6270877 & 2.0662947 & 17.33 & 17.18 & 17.06 & 16.95 &   -   & CR47 \\ 
BSS-37 & 229.6244175 & 2.0693612 & 16.89 & 16.41 & 16.51 & 15.71 &   -   & CR46 \\ 
BSS-38 & 229.6180419 & 2.0724090 & 17.37 & 17.23 & 17.12 & 17.00 &   -   & CR34 \\ 
BSS-39 & 229.6311963 & 2.0857800 & 18.31 & 17.33 & 17.40 & 16.76 &   -   &  -   \\ 
BSS-40 & 229.6297499 & 2.0664961 & 18.16 & 17.58 &   -   & 17.27 &   -   & CR76 \\ 
BSS-41 & 229.6443367 & 2.0872809 &   -   &   -   & 17.50 & 17.23 &   -   & CR9 \\ 
BSS-42 & 229.6448646 & 2.0738335 &   -   &   -   & 16.53 & 16.06 & 15.95 & CR32 \\ 
BSS-43 & 229.6460645 & 2.0748695 &   -   &   -   & 16.64 & 16.44 & 16.66 & CR30 \\ 
BSS-44 & 229.6481631 & 2.0718829 &   -   &   -   & 16.72 & 16.61 & 16.87 & CR37 \\ 
BSS-45 & 229.6433942 & 2.0760163 &   -   &   -   & 17.03 & 16.79 & 16.91 & CR28 \\ 
BSS-46 & 229.6439884 & 2.0775670 &   -   &   -   & 17.44 & 16.99 & 16.81 &  -   \\ 
BSS-47 & 229.6180420 & 2.0598328 &   -   &   -   &   -   & 17.18 & 17.12 &  -   \\ 
BSS-48 & 229.6092873 & 2.1680914 &   -   &   -   & 16.85 & 16.68 &   -   & OR2 \\ 
BSS-49 & 229.6723094 & 2.0882827 &   -   &   -   & 16.94 & 16.64 &   -   & OR9 \\ 
BSS-50 & 229.6006551 & 2.0814678 &   -   &   -   & 17.00 & 16.74 &   -   & OR10 \\ 
BSS-51 & 229.6669956 & 1.9781808 &   -   &   -   & 17.20 & 16.74 &   -   & OR1 \\ 
BSS-52 & 229.5949935 & 2.0469325 &   -   &   -   & 17.69 & 17.46 &   -   & OR4 \\ 
BSS-53 & 229.6706625 & 2.0695464 &   -   &   -   & 17.82 & 17.50 &   -   &  -   \\ 
BSS-54 & 229.6667908 & 2.1149550 &   -   &   -   & 17.82 & 17.72 &   -   &  -   \\ 
BSS-55 & 229.7370667 & 2.0323392 &   -   &   -   & 17.80 & 17.42 &   -   & OR23 \\ 
BSS-56 & 229.5476990 & 2.0112610 &   -   &   -   & 16.88 & 16.60 &   -   & OR5 \\ 
BSS-57 & 229.6711255 & 1.9415566 &   -   &   -   & 16.98 & 16.64 &   -   &  -   \\ 
BSS-58 & 229.4381714 & 2.0302088 &   -   &   -   & 17.75 & 17.33 &   -   &  -   \\ 
BSS-59 & 229.7408412 & 2.3399166 &   -   &   -   & 17.49 & 17.08 &   -   &  -   \\ 
BSS-60 & 229.3218200 & 2.3271022 &   -   &   -   & 16.34 & 16.09 &   -   &  -   \\ 
\hline
\enddata
\tablecomments{The first 41 BSS have been identified in the WFPC2 sample;
BSS-42--46 are from the complementary ACS observations; BSS-47--59 are from
the WFI data-set.  BSS-59 lies beyond the cluster tidal radius, at $\sim
24\arcmin$ from the center.  The last column list the corresponding BSS in
W06 sample, with "CR" indicating their "Core BSS" and "OR" their "Outer
Region BSS".}
\label{tab:BSS}
\end{deluxetable}

\begin{deluxetable}{rrrcc}
\tablecaption{Number counts of BSS and HB stars}
\scriptsize
\tablewidth{7.5cm}
\startdata \\
\hline \hline
  $r_i\arcsec$  &    $r_e\arcsec$ & $N_{\rm BSS}$ & $N_{\rm HB}$ & $L^{\rm samp}/L_{\rm tot}^{\rm samp}$ \\
\hline 
  0 &   27 &  22 & 94 &  0.14 \\
  27 &  50 &  15 & 94 &  0.16 \\
  50 & 115 &  10 &135 &  0.26 \\
 115 & 150 &   3 & 46 &  0.09 \\
 150 & 210 &   2 & 52 &  0.10 \\
 210 & 300 &   0 & 45$^\dag$ &  0.10 \\
 300 & 450 &   4 & 42$^\dag$ &  0.09 \\
 450 & 800 &   2 & 38$^\dag$ &  0.06 \\
\hline 
\enddata
\tablecomments{$^\dag$~The $N_{\rm HB}$ values listed here are those
corrected for field contamination (i.e., 1, 2 and 8 stars have been
subtracted to the observed number counts in these three external annuli,
respectively).}
\label{tab:annuli}
\end{deluxetable}

\end{document}